\documentstyle[referee]{laa}     
\input{psfig}

\begin{document}

\def\etal{{\it et al.\ }}
\def\vx{\hbox{\bf x}}
\newcommand{\asb}[3]{#1, Ann. sco. sci. Bruxelles, #2\rm, #3}
\newcommand{\nat}[3]{#1,  Nature, #2\rm, #3}
\newcommand{\np}[3]{#1,  Nucl. Phys.,  #2\rm, #3}
\newcommand{\apj}[3]{#1,  ApJ,  #2\rm, #3}
\newcommand{\aj}[3]{#1,  AJ,  #2\rm, #3}
\newcommand{\apjl}[3]{#1, ApJ,  #2\rm, L#3}
\newcommand{\apjsup}[3]{#1,  ApJS, #2\rm, #3}
\newcommand{\aeta}[3]{#1, A\&A,  #2\rm, #3}
\newcommand{\annrev}[3]{#1, ARA\&A,  #2\rm, #3}
\newcommand{\mnras}[3]{#1,  MNRAS, #2\rm, #3}
\newcommand{\jrasc}[3]{#1,  JRASC,  #2\rm, #3}
\newcommand{\physlettb}[3]{#1, Phys. Lett. B, #2\rm, #3}
\newcommand{\physrevlett}[3]{#1, Phys. Rev. Lett., #2\rm, #3}

\thesaurus{02        
          (03.12.1  
           )}
\title{X--Ray galaxy clusters: constraints on models of 
galaxy formation}
\author{J. Oukbir$^1$, J.G. Bartlett$^2$ 
\and A. Blanchard$^2$}
\offprints{J. Oukbir}
\institute{ $^1$CE--Saclay DSM/DAPNIA/SAp, Orme des Merisiers, F--91191 
Gif--sur--Yvette Cedex\\
$^2$Observatoire Astronomique de Strasbourg, 11 rue de l'Universit\'e,
F--67000 Strasbourg}
\maketitle
\newpage

\begin{abstract}
We present a self--consistent approach to the modeling of X--ray
clusters of galaxies in a flat universe.  
Employing the Press \& Schechter (1974) formalism to
derive the mass function and
relating the observable 
properties of clusters to their virial mass allows us 
to study the cluster X--ray distribution functions and their evolution
with redshift. This approach 
differs from some results based on the X--ray luminosity
function, which we argue is subject to modeling uncertainties.
We 
obtain stringent constraints on the power spectrum of the initial
density perturbations assumed to seed galaxy formation by comparing
the theoretical temperature function with available data: 
the amplitude $\sigma_8$ is found to be $0.57 \pm  0.05$, while 
the local spectral index $n$ is falling in the range
$-2.4 \leq n \leq -1.5$ on scales between 5 and 15 
${\mbox {\rm h}}^{-1}$Mpc:
the standard CDM model
is clearly ruled out, independently of the normalization.
We further examine the evolutionary properties of X--ray clusters.
Our approach greatly clarifies the situation: contrary to 
some previous claims, we find that flat models 
are in reasonable agreement with  
the observed number of clusters at high redshifts.
We also examine the contribution of clusters to the X--ray
counts and to the soft X--ray background and compare the expected 
values in the case of the $\Omega_0=1$ and $\Omega_0=0.2$
universes. The number counts are in agreement with the observations, 
further confirming the relevance of our modeling. We 
conclude that although clusters are not the primary source of the
soft X--ray background, their contribution is nevertheless
non--negligible.
This is particularly important for they could escape the 
constraints imposed on possible point sources contributing
to the background.
Finally, we briefly examine the case of non--Gaussian fluctuations
and point out the degeneracy between the value of the power spectrum 
index and the nature of the fluctuations, 
that these models imply.

\end{abstract}
\newpage

\section{Introduction}

      	The ``standard'' picture for the origin of the large--scale
distribution of matter in the universe is based on the gravitational
growth of initially small density perturbations assumed to be present
from the very earliest moments of cosmic time (Lema\^{\i}tre 1933).
This idea has received considerable attention from theorists and 
recently some spectacular observational support: the detection of 
temperature fluctuations in the Cosmic Background Radiation
(CBR) by the COBE instrument DMR (Smoot {\etal} 1992). To this
discovery one may today add a host of other claimed 
detections (see, for example, Scott {\etal} 1995).  
In the gravitational instability scenario, the density
field can be  quantitatively described by the power spectrum of the
perturbations and by their higher order moments.  In this paper,
although the non--gaussian case will be briefly discussed, we 
will be primarily concerned with Gaussian perturbations 
for which the power spectrum provides a sufficient description. 
The uncertain origin of the density perturbations translates
into an inability to calculate from first principles the form
of the power spectrum, and the lack of a specific value
for the density parameter exacerbates the problem.
The theory of inflation alleviates this theoretical uncertainty
by predicting that the primordial spectrum follows
a power--law: $P(k)\propto k^{n_p}$ with $n_p \le 1$, the exact value 
of $n_p$ depending upon the possible existence of gravitational 
waves (Lucchin {\etal} 1992).  In addition, inflation sets the cosmic density 
to the critical value ($\rho_c = 3H_0^2/8\pi G$, where 
$H_0 = {\mbox {\rm h}}100 
\ \mbox{\rm km}/\mbox{\rm s}/\mbox{\rm Mpc}$ is the current value of the 
Hubble constant).
Once the nature of the (necessarily non--baryonic) dark matter 
is specified, one may calculate the final, evolved power spectrum
as a function of the amplitude and index $n_p$ of the initial spectrum.
This is the spectrum relevant to galaxy formation.
 
	Among the possible scenarios based on inflation,  
the cold dark matter (CDM) model has proven very successful in 
explaining many observed properties of the universe on scales ranging
from galaxies to galaxy clusters, i.e. between a few tens of kpc 
and a few Mpc (for a review on the subject, see Frenk 1991).
The amplitude of the power spectrum is the only 
free parameter of the ``standard'' model: one adopts $n_p=1$.

	The galaxy cluster population provides some of the most stringent 
constraints on models of galaxy formation, essentially because 
clusters are rare objects and, hence, their properties are sensitive 
to the underlying density fluctuations.  The obviously important role 
of baryons in the determination of the observed properties of clusters 
would seem to necessitate hydrodynamical simulations in order
to derive any such constraints.  Evrard (1989) has pioneered this 
interesting and important approach.  Nevertheless, the ensemble 
properties of clusters, like their optical and X--ray 
luminosity functions, or their velocity and temperature distribution 
functions, are difficult to address directly by numerical simulations 
because the size of the numerical box must be very large in order 
to contain a sufficient number of clusters; an analytical approach
remains an effective alternative.

	Kaiser (1986) made an important theoretical step in this 
direction by proposing and then using simple scaling laws for cluster 
properties to derive the evolution of the ensemble properties.
The corroboration of the Press \& Schechter (1974) 
(PS) mass function 
by some recent numerical simulations (Efstathiou {\etal} 1988,
Carlberg \& Couchman 1989, Gelb \& Bertschinger 1994, Eke {\etal} 1996; 
see Brainerd \& Villumsen 
1992 for an alternate point of view) provides us with
an even more powerful tool for constraining galaxy formation 
theories: the simple PS formula gives us the mass function of 
structures at any redshift for any theory of initially Gaussian 
fluctuations.  Provided that the relation between 
some observed quantity (for instance, the luminosity) and the mass is known,
perhaps provided by hydrodynamical simulations of a small number of
clusters, both the cluster ensemble properties (e.g. the luminosity 
function) and their evolution can be predicted and compared to 
observations.  For example, Schaeffer and Silk (1988) showed
that the CDM scenario reproduces well the optical luminosity
function of galaxy clusters. Evrard (1989), 
using the number density of galaxy clusters with velocity 
dispersions greater than 
$1350\,\mbox{\rm km}\,\mbox{\rm s}^{-1}$ at redshifts 
lower than $0.1$, concluded
that the bias parameter $b$ -- which is defined to be the inverse
of the rms value of mass fluctuations within spheres of 
radius $8{\mbox{\rm h}}^{-1}{\mbox{\rm Mpc}}$ --
must be of the order of $1.5$, 
inconsistent
with the higher values of $b$ required to explain galactic properties.
He pointed out that an even smaller 
value of the bias is necessary to explain
the three high velocity dispersion clusters listed by Gunn (1989) at 
redshifts greater than 0.1.  Peebles {\etal} (1989) reached a similar
conclusion by using a variety of present day cluster properties.  
However, Frenk {\etal} (1990) argued, by constructing artificial 
cluster catalogues from numerical simulations, that the cluster 
velocity dispersion is not a property from which 
reliable constraints can be derived on models  because 
projection effects along the line
of sight can contaminate the galaxy samples and significantly increase
the estimated velocity dispersion. 
Since then, the standard CDM model has met with some further serious 
problems, for example an inability to explain the angular
correlations of galaxies detected with the Automatic Plate Measuring
Machine (APM) (Maddox {\etal} 1990).
This has shed doubt on the validity of ``standard'' CDM.
In addition, the amplitude indicated by the 
COBE temperature fluctuations, corresponding to $b\sim 1$ rather than
the advocated $b\sim 2$, is generally considered too high for 
the model to be viable (see, however, Bartlett \& Blanchard 1994, 1996), 
although some other authors have suggested that this high normalization 
can explain the observations once non--linear effects have been properly 
accounted for (Couchman \& Carlberg 1992).  

	To solve the various problems
faced by the standard version of CDM, changes to the power spectrum 
have been proposed, 
such as suppressing the small--scale power by mixing in a small 
amount of hot dark matter (Bond {\etal} 1980, Bond \& Szalay 1983,
Dekel 1984, Schaefer \& Shafi 1992, Davis {\etal} 1992) or by altering 
the primordial value of $n_p$ (Cen {\etal} 1992, Cen \& Ostriker 1993) 
(so--called ``tilted'' CDM models). A list of further other possibilities is
given by McNally \& Peacock (1995). All of this leads us to 
reconsider the form of the power spectrum and adopt the point
of view that it is an unknown which we wish to constrain.  
This analysis constitutes an alternative to a direct analysis of
galaxy surveys (Peacock \& Dodds 1994) 
for which the amplitude of the bias is unknown
and does not permit direct access to the mass distribution.
In particular, we will use the cluster population for this 
purpose: we will assume that over cluster scales  
even the evolved power spectrum can be approximated by a power--law,
and then we will use the ensemble cluster properties to place limits
on the amplitude and spectral index $n$.  
The adoption of a power--law
is not really restrictive as most currently 
considered models lend themselves to this approximation (this may 
not necessarily be the case in purely baryonic models, in which
the Jeans mass may strongly influence the perturbations on cluster
scales and in a manner dependent upon the ionization history).

	In principle, both optical and X--ray data can be used to constrain
models, but, as emphasized by Frenk {\etal} (1990), the optical 
properties are subject to projection effects.  If indeed important, 
such effects can alter both the optical luminosity   
and the velocity distribution functions.  Moreover, the relationship 
between the overall mass of a cluster and its constituent galaxies 
could very well be complicated by the 
non--linear physics of galaxy formation (Evrard {\etal} 1994).  
The X--ray properties of a cluster offer
an interesting alternative as they should not suffer the same 
severe projection effects.  However, the observed X--ray 
luminosity of clusters is dominated by their
core radius, and the physical origin of this core is unknown.
This makes it difficult to relate the X--ray luminosity to the
cluster mass, a point we will discuss in greater detail below
and which will lead us to focus on the temperature distribution 
function.

  	In recent years many authors have calculated the 
ensemble properties of X--ray clusters expected in various scenarios (Henry 
\& Arnaud 1991, Blanchard \& Silk 1991, Kaiser 1991, Pierre 1991, 
Lilje 1992, Oukbir \& Blanchard 1992, Bahcall \& Cen 1993,
Bartlett \& Silk 1993, Blanchard {\it et al.} 1994, 
Colafrancesco \& Vittorio 1994, Balland \& Blanchard 1995
Liddle {\etal} 1995, Eke {\etal} 1996)
and compared the results with observations
(Edge {\etal} 1990, Henry \& Arnaud 1991) in order 
to derive constraints on the power spectrum.
Henry \& Arnaud (1991) found that the spectral index $n$ and the bias
$b$ of the density perturbations must be $-2.1$ 
and $1.7$, respectively, to reproduce
their observed temperature distribution function. 
Blanchard \& Silk (1991) claimed
that the CDM model is marginally consistent with the Edge {\etal} (1990)
data if the bias parameter is close to $1.5$, but that $n=-2$ 
with $b \approx 1.7$ is
favored over the CDM value of $n\approx -1$ on cluster scales. 
However, Kaiser (1991) argues that the observed 
evolution of the luminosity function needs an index 
closer to $-1$. Lilje (1992) has shown that $\Omega_0=0.2$, 
flat CDM models
need to be antibiased in order to reproduce the temperature distribution
function.  He also noticed that at high redshifts the temperature 
distribution evolves differently depending on the value
of $\lambda_0$.  Both of these conclusions are consistent with
the results of Bartlett \& Silk (1993). 
Oukbir \& Blanchard (1992, 1996) have shown that an unbiased open 
universe with $\Omega_0=0.2$ is compatible with the observed temperature 
distribution and that the redshift distribution of X--ray clusters 
is a powerful test of the mean density of the universe.
Colafrancesco \& Vittorio (1994) investigated the
constraints imposed by the cluster luminosity function on a 
variety of models normalized to COBE, extending the analysis 
of Bartlett \& Silk (1993), although reaching quite different 
conclusions. 

Given this large list of different analyses, it would 
seem difficult to derive a consistent set of constraints from observations
of the cluster population.  In this paper we re--examine the modeling 
of X--ray clusters to clarify the situation.  We construct a self--consistent 
set of relations between observable cluster properties and the 
theoretically relevant virial mass.  We then use these
relations to obtain robust constraints on the power spectrum
(i.e. on the amplitude and spectral index). 
These constraints are applicable on scales from 5 to 15 
${\mbox {\rm h}}^{-1}$Mpc.  
In this work we consider only a flat, hierarchical, dark matter dominated 
universe with $\Omega_0= 1$.   
We start in the next section with a presentation 
of the arguments supporting the PS formula.  
In the following section, we discuss the observed properties of individual 
clusters and then relate them to the mass appearing in the theoretical 
mass function (Sect. 3).  In the fourth section we derive the 
power spectrum parameters which best reproduce the observed 
temperature distribution function.  
In the fifth section we dicuss the expected evolution of
the luminosity function and compare the model to the 
high redshift observations. The sixth section presents
predictions for the cluster number counts as a function of X--ray
flux and an estimate of the cluster contribution to the soft 
X--ray background.  Finally, the seventh section contains a brief 
discussion of non--Gaussian fluctuations. 
In the last section we summarize our results.

\section{Theoretical mass function}

In the gravitational instability scenario, virialized objects like
galaxy clusters form from initially small density fluctuations
that grow under the influence of gravity. The density field is specified
by its power spectrum and the statistical nature of the fluctuations. 
It is generally assumed that the field is Gaussian, although some
consequences of non--Gaussianity have been examined 
(Weinberg and Cole 1992), a question to which we return in Sect. 7.  
For the power spectrum, we adopt
a simple power law,
$$P(k)\propto k^n, $$
over the mass range corresponding 
to galaxy clusters.  
In hierarchical models, such as CDM, the variance of the  
density field diverges on small scales,
and so one must work with a smoothed version (see, for instance, 
Bardeen {\etal} 1986).  The variance of the density 
field smoothed with some 
window function $W$ on a scale corresponding to mass 
$M$ is 
$$\sigma^2 (M)= {2\over (2\pi)^2}\int_0^{\infty}dk\,k^2P(k)
\tilde{W}^2(k\,R), $$
where $\tilde{W}$ is the Fourier transform of the window function. 
Davis $\&$ Peebles (1983) found that the variance
of galaxy counts within spheres of radius 
$8{\mbox {\rm h}}^{-1}\mbox{\rm Mpc}$, 
quoted as 
$\sigma_{{\mbox{\rm{\tiny gal}}}}
(8{\mbox {\rm h}} ^{-1}\mbox{\rm Mpc})$,
is close to one. If the galaxy distribution
follows the mass distribution, then the variance of the density 
perturbations on the same scale would  also be 
equal to unity.
However, if the galaxy distribution is biased relative to the mass
distribution, then 
the variance of mass fluctuations in a sphere of radius $R$
containing a mass $M = 4/3 \pi \rho R^3$  can be written in the following
way:
\begin{equation}
\label{eq.sigma}
\sigma(M) = {1\over b}\left({M_8\over M}\right)^{({n+3})/6},
\end{equation}
Accordingly, in the following, the value of $b$ will correspond 
to $1/\sigma_8$ where $b$ is the bias parameter.  
The value of $b$ advocated to explain the observed abundance of 
galaxies, their correlations and their velocity dispersion was 
in the range $2-2.5$.
This large value of $b$ met with difficulty 
in other quarters (Valls--Gabaud {\etal} 1989).

Since in one dimension the early stage 
of the non--linear collapse
is entirely determined by the amplitude of the local mean density,
the exact solution for the collapse of an overdensity
is calculable until the first orbit crossing occurs. 
In three dimensions, the solution is also known for 
the case of a spherical matter distribution (Lema\^{\i}tre 1933,
Gunn \& Gott 1972, Peebles 1980).
This so--called spherical model
can be used to model the non--linear collapse of a cluster, 
which one then finds is driven by the value of the density field 
smoothed with a top--hat window of size comparable to that of the cluster. 
It is well known that when the linear density field 
reaches the value 1.68, the density becomes singular for a purely
spherical collapse.  In reality, the spherical symmetry is broken by
the development of substructures 
which instead leads to the formation of a stationary state,  
the ``virial'' equilibrium.
The final radius of the collapsed object is expected to be 
half of its maximum expansion radius, corresponding to a
density contrast of the order of 200.
In the absence of significant fragmentation during the collapse, 
initial density fluctuations of the field smoothed on the scale $R$
are expected to collapse to structures with a typical mass 
$M = 4/3\pi \rho R^3$, where $\rho$ is the mean cosmological 
background density. Within this framework, it is in principle possible 
to relate the ensemble characteristics of non--linear objects to the 
statistical properties of the initial density field.

In practice one deals with the initial  density field linearly 
extrapolated to the present
epoch $z=0$. For instance, the rms fluctuation on some scale $M$ is:
$$\sigma_0 (M) =  {D(z=0)\over D(z_i)}\sigma_i (M). $$
In this relation $D$ is the growing mode solution 
of the linearized growth equation
and $z_i$ is the redshift corresponding to some
early time at which the fluctuations in the universe were still linear.
For the case of $\Omega_0=1$ and a vanishing cosmological constant,
$\lambda_0=0$, $D\propto a$, where $a$ is the expansion factor.

Despite the fact that the spherical top--hat model permits a considerable 
simplification of the actual development of non--linearities, 
the precise calculation of the number density of collapsed 
objects of mass $M$ remains an extremely 
complicated problem.
One major difficulty comes from the fact that a given region of space
identified as non--linear on some scale might in fact  form
part of a still larger non--linear structure.  This is the so--called
``cloud--in--cloud'' problem.  However, one may assume that, being rare,
massive objects, such as galaxy clusters, originate from nearly
isolated density fluctuations, for which the cloud--in--cloud effect 
should be less important. In this case, the spherical model
is likely to be a good description of the nonlinear evolution. 
Indeed, Bernardeau (1994)
has shown that the rare density
fluctuations of a Gaussian random field follow exactly the dynamics
of the spherical model.

Using the spherical model, Press \& Schechter (1974) proposed a derivation
of the mass function of virialized objects. They argued that 
the fraction of 
matter in the form of non--linear objects with mass greater than $M$ 
may be evaluated by:
$$
\int_{\nu_t}^{\infty}\frac{1}{\sqrt{2\pi}} \exp (-\nu^2/2)d\nu,
$$
where $\nu_t$ is the threshold for non--linear collapse, which in the spherical
model is given by
$$
\nu_t = \frac{\delta_{c,0}(z)}{\sigma_0(M)}
$$
For an Einstein--de Sitter universe, $\delta_{c,0} = 1.68(1+z)$.
The mass distribution function is then easily derived:
$$N(M,z) dM = - \sqrt{{1\over 2\pi}}{\rho_b\over M}
{\delta_{c,0}(z)\over \sigma_0^2}
 {d\sigma_0\over dM}            
\mbox{\rm exp}\left(-{\delta_{c,0}^2\over 2\sigma_0^2}\right)dM.$$

Unfortunately, this result predicts that only 
half of the mass of the universe ends up in virialized
objects, whereas one expects that at each epoch in a hierarchical scenario
all of the mass should be bound in virialized objects.
Press \& Schechter (1974) side--stepped this problem by
arbitrarily multiplying this mass function by a factor of 2.
Several authors have recently re--examined the issue and proposed
solutions (Peacock \& Heavens 1990, Cole 1991, Blanchard {\etal} 1992a).

The problem has also been considered by Bond {\etal} (1991). 
Instead of smoothing
the density contrast $\delta(x)$ with a filter $W(x,R)$, they
adopted a low--pass, top--hat filter in Fourier space. 
At a given point $x$ in real space,
the value $\delta_s$ of the smoothed density contrast executes a Gaussian
random walk as the size of the filter is increased from $k_c$ to
$k_c + \Delta k_c$ (corresponding to a decrease in the size
of the window in real space).  The step size of the random walk is 
a Gaussian random variable whose variance depends only
on the power spectrum $P(k)$ and on the value of $k_c$ (all of this
applies only for Gaussian density fields).  By 
identifying points which first pass the critical density $\delta_c$ 
at a particular value $k_c=K_c$ as objects of mass 
$M = 6\pi^2\rho_b a^3 K_c^{-3}$, 
the authors were able to recover the PS formula, 
including the troublesome factor of 2.

On the other hand, Blanchard {\etal} (1992a) 
point out that with the usually adopted 
simplifications (i.e. the collapse being driven essentially by the mean 
local density and the neglect of fragmentation), 
it is always possible to write 
the mass function in an exact way as:

\begin{equation}
\label{eq.MF}
N(M,z) dM = - 
{\rho_b\over M}
{\delta_{c,0}(z)\over \sigma_0^2}
 {d\sigma_0\over dM}            
F\left({\delta_{c,0}^2\over 2\sigma_0^2}\right)dM,
\end{equation}

where $F(\nu)$ represents the fraction of volume covered by non--linear spheres
of radius greater than $R$ (the smoothing scale corresponding 
to mass $M$).
Although the function $F$ is well defined, its calculation
represents an extremely complicated statistical problem 
for Gaussian random fields.
It would appear, then, that the normalization problem 
of the PS formula is due to the assumption that only
points which are at the center of non--linear spheres of radius $R$
are counted, rather than all points (not just the centers) residing
in spheres of radius $R$ or greater. 

We justify our use of the PS formalism by its
surprisingly good fit to the mass functions found in numerical simulations.
This was first emphasized by Efstathiou {\etal} (1988), who examined
the cluster multiplicity function for different values of $n$.
Let us consider the more recent simulations of White {\etal} (1993).
With an error of less than 50\% in the mass, 
the simulation distributions agree with the PS formula for masses above
$5\times 10^{13}\mbox{\rm M}_{\odot}$, where the abundance is
$10^{-7} {{\mbox {\rm h}}^{-1}\mbox{\rm Mpc}}^{-3}$. 
We may take this to mean that
the PS mass function is reliable 
into the regime where it accounts for $\approx 10^{-4}$ of the total mass.
Roughly speaking, real clusters with a mass of a couple of $10^{15}
\mbox{\rm M}_{\odot}$ 
have an abundance of a few $10^{-8} 
({\mbox {\rm h}}^{-1}\mbox{\rm Mpc})^{-3}$ and
thus represent about $10^{-4}$ of the total mass.  For clusters with $T\approx 
14$ keV this means an abundance of  $10^{-9} 
({\mbox {\rm h}}^{-1}\mbox{\rm Mpc})^{-3}$.
We conclude
that the PS formula can be applied to confidently predict  
the abundance of X--ray 
clusters over the full range of observed temperatures.

\section{The $T-M$ and $L_X-M$ relations}

\subsection{The $T-M$ relation}
We cannot directly relate the mass function to observations because we have
very little information on the actual virial mass of clusters.
Note that in the spherical top--hat model, the virial radius is
$$ R_V = 1.69 {\mbox {\rm h}}^{1/3}M_{15}^{1/3} (1+z)^{-1} 
{\mbox {\rm h}}^{-1}\mbox{\rm Mpc}, $$
which extends out beyond the region of currently available mass
determinations.  
In order to obtain a fruitful comparison of the 
theoretical mass function with the observations, it is therefore
necessary to construct trustworthy relations between X-ray 
properties and cluster virial masses.  This ``virial'' mass
should be understood in the sense in which it is employed in 
the mass function:  it is the mass contained within the region of mean 
contrast density $\sim 200$.  Here we consider the $T-M$ and 
$L_X-M$ relations.  We shall see that the former is more reliable.

 	Spectroscopic studies have demonstrated 
that the X-ray emission is produced by thermal bremsstrahlung in 
an hot, optically thin intracluster plasma with a temperature of 
approximately $10^8\ K$.  The detailed history of this gas is not
well known.  The presence of the $7\ \mbox{\rm keV}$ iron emission line 
indicates that the intracluster medium (ICM) has been 
partially processed through the stars of the cluster galaxies.  
However, the large mass of the ICM, typically several times 
greater than the cluster stellar mass, leads one to believe 
that the majority of the gas is primordial in origin, since 
it seems difficult that the galaxies lose through winds 
or ram pressure stripping more than 50\% 
of their initial mass. In addition,
the measured metallicities close to third $ Z_{\odot}$ can 
be accounted for by a bimodial star formation model
(Arnaud {\etal} 1992a).

   We will work under the hypothesis that the cluster gas
is in hydrostatic equilibrium with an isothermal temperature
profile. 
Despite the lack of rigorous evidence, the latter 
point is at least consistent with the majority of data. Under 
these conditions we may write 
$$ {kT\over \mu m_p}{d\mbox{\rm ln}\rho_{\mbox{\rm {\tiny gas}}}(r)
\over d\mbox{\rm ln}r}
= -{GM(r)\over r},$$
where $kT$ and $\rho_{\mbox{\rm {\tiny gas}}}$ are, 
respectively, the temperature and 
the density of the gas, $\mu$ is the mean molecular weight, $m_p$ is 
the proton mass and $M(r)$
is the binding mass of the cluster. 
The observed X--ray surface brightness profile can be directly
converted to a three dimensional density profile:
$$\rho_{{\mbox{\rm{\tiny gas}}}}(r) 
= \rho_{0,{\mbox{\rm{\tiny gas}}}}\bigl(1
+{(r/r_{\mbox{\tiny c},{\mbox{\rm{\tiny gas}}}})}^2
\bigr)^{-3\beta_{\mbox{\rm{\tiny fit}}}/2}.
$$
This is just the deprojection of the isothermal $\beta$ form known 
to fit the surface brightness profiles of clusters.
Best fit values for $\beta_{\mbox{\rm {\tiny fit}}}$ 
are typically around 0.6 (Jones \& Forman 1984).
Using this value, the relation
between the temperature and the virial mass can be evaluated as
$$ T = 2\frac{\mu m_p}{k}\frac{GM}{R}$$
This result is in
reasonable agreement with the hydrodynamic simulations 
of Evrard (1990a, 1990b), although he obtained a constant of
proportionality which is about 20 \% lower. 
Taking this into account, the relation between mass and temperature becomes
$$kT = 4  M_{15}^{2/3}(1+z){\mbox {\rm h}}^{2/3}\,\mbox{\rm keV}$$
In this expression
$M_{15}$ is the cluster virial mass in units of 
$10^{15} \mbox{\rm M}_{\odot}$. 
More recently, Evrard {\etal} (1996) showed 
that in the CDM case, this relation between mass and temperature holds 
with a very good accuray.
This equation will allow us to transform the mass function into a 
temperature function which we can compare to observations.  

The temperature
can also be related to the initial, comoving radius containing the mass:
$$kT = 4\,\mbox{\rm keV} \left(\frac{R}
{8{\mbox {\rm h}}^{-1}\mbox{\rm Mpc}}\right)^{2/3}(1+z).$$
This last relation illustrates an important point: as the temperature is 
independent of
the Hubble constant when scales are expressed in 
${\mbox {\rm h}}^{-1}$Mpc, 
a model corresponding to a given power spectrum and normalized to 
a scale also measured in ${\mbox {\rm h}}^{-1}$Mpc 
(such as $\sigma(M_8)$) will
produce the same cluster abundance per 
$({{\mbox {\rm h}}^{-1}\mbox{\rm Mpc}})^{-3}$. Therefore, 
the constraints on $n$ and $b$ inferred from the observed abundances of 
clusters are independent of the value 
of the Hubble constant.

\subsection{The $L_X-M$ relation}
	The bolometric X--ray luminosity of a galaxy cluster due to thermal
bremsstrahlung depends strongly on the gas density profile:
$$L_X \propto \int_0^{\infty} \rho_{\mbox{\rm {\tiny gas}}}^2(r) 
T^{1/2} 4\pi r^2 dr. $$
Usually, one assumes that the ICM represents a constant fraction $f_g$ 
of the cluster virial mass.
Adopting this hypothesis and assuming an identical radial distribution for
both the gas and the dynamical mass, Kaiser (1986) derived a scaling law
for the X--ray luminosity: $L_X\propto M^{4/3} (1+z)^{3.5} $.
However, there are reasons to suspect this scaling law. 
As emphasized by Blanchard {\it et al.} (1992b), 
the total luminosity of a cluster is 
dominated by the mass of the gas core and the
self--similar scaling applies only if
the mass of the gas core scales as the virial mass.
However, the formation of a core in the gas distribution is not 
well understood and may result from any of several 
physical processes, including cooling in the
center of the cluster or in the smaller structures from which the cluster was 
built, preheating by a first generation of collapsed objects, or
gas ejection from galaxies.  Thus, theoretical modeling of the X--ray 
luminosity is dangerously uncertain. Indeed, if one combines the 
self--similar $L_X-M$ relation with the highly reliable theoretical 
$T-M$ relation, then one obtains a $L_X-T$ correlation whose 
shape 
is in severe conflict with local data.
In addition, Blanchard \& Silk (1991) 
and Evrard $\&$ Henry (1991)
have shown that self--similar scaling produces
a luminosity function which disagrees with observations for both 
the CDM model
and for a model with a power--law power spectrum with index $n=-2$.
An alternative to the self--similar scheme
is to parametrize the luminosity--mass relation as a power law of the cluster
mass and redshift, $L_X \propto L_0 M^p (1+z)^s$, and to consider values
of the free parameters that fit the observed luminosity function.  But as
these authors have noted, there is an observational degeneracy
between the shape of the assumed initial power spectrum and the chosen
$L_X-M$ relation.  
Perhaps the best way to deduce the true
$L_X-M$ relation is through the observed $L_X-T$ relation. 
This seems a more trustworthy approach if one believes that the 
temperature reflects the virial energy of
the cluster.  In the remainder of this paper, we follow this procedure 
and use the local $L_X-T$ relation. 
Edge \& Stewart (1991) have found that
$ L_X = (10^{43.05}\mbox{\rm erg}\,\mbox{\rm s}^{-1})
T_{\mbox {\rm {\tiny keV}}}^{2.62}$
using EXOSAT data.  This relation is very close to that found by 
Henry \& Arnaud (1991) using a compilation of data from EXOSAT, 
HEAO/OSO and the Einstein satellite.  Additionally, Edge \& Stewart (1991) 
have also given the correlation
$T = (10^{-12.73}\,\mbox{\rm keV})
L_{X,\mbox{\rm {\tiny erg}}\,\mbox{\rm {\tiny s}}^{-1}}^{0.30}$ 
resulting from a minimization of the residuals in $\mbox{\rm log} T$.
Using the result of a double regression fit in which the slope 
is defined as
the square--root of the product of the two individual regression slopes, 
we find $L_X\propto T^3$.
We normalize this relation at $7\,\mbox{\rm keV}$, the temperature at which
the two regressions cross each other, to finally obtain

\begin{equation}
\label{eq.LT}
L_X = (10^{42.7}\,\mbox{\rm erg}\,\mbox{\rm s}^{-1})
T_{\mbox{ \rm {\tiny keV}}}^{3}.
\end{equation}

This expression represents the bolometric X--ray luminosity and 
we must correct for the fraction
$$f_{\mbox{\rm {\tiny band}}}(z) 
=  \int_{E_1(1+z)}^{E_2(1+z)} dE\, e^{-E/kT}/kT$$
actually collected in the relevant energy band [$E_1$--$E_2$].  

\section{Constraints on $n$ and $b$}

As we have just emphasized, the relation 
between the ICM temperature and the cluster virial mass
is, in contrast to the luminosity--mass relation, relatively well understood.  
For this reason, we prefer in the 
following to use the temperature distribution function rather than
the luminosity function to draw our conclusions on the power spectrum
of density fluctuations.

	The two existing cluster temperature distribution
functions were derived from X--ray all--sky surveys.
The Edge {\etal} (1990) temperature function was 
derived from $55$ clusters with
fluxes greater than  $1.7\times 10^{-11}\mbox{\rm erg}\,\mbox{\rm cm}^{-2} 
\mbox{\rm s}^{-1}$ in the 2--10 keV band. 
It was constructed by correcting the Piccinotti 
{\etal} (1982) survey  (which is supposed to be complete down to 
$3.1\times 10^{-11}\mbox{\rm erg}\,\mbox{\rm cm}^{-2} \mbox{\rm s}^{-1}$), 
for both mis--identification and confusion, and then by including
clusters with fluxes greater than
$1.7\times 10^{-11}\mbox{\rm erg}\,\mbox{\rm cm}^{-2} \mbox{\rm s}^{-1}$.
  The supposed clusters with insufficient information were 
excluded from the analysis,
but the authors estimated from the $\mbox{\rm log}N$--$\mbox{\rm log}S$ 
and $V/V_{\mbox{\rm {\tiny max}}}$ 
distributions that their sample is $100\%$ complete down to 
$3.1\times 10^{-11}\mbox{\rm erg}\,\mbox{\rm cm}^{-2} \mbox{\rm s}^{-1}$  
and $70-90\%$ complete down to $1.7\times 10^{-11} \mbox{\rm erg}\,
\mbox{\rm cm}^{-2} \mbox{\rm s}^{-1}$.

Henry \& Arnaud (1991) also used the Piccinotti {\etal} (1982)
all sky sample with corrections for source confusion to derive the 
temperature distribution function. 
They find that a power--law with index similar to that derived 
by Edge {\etal} (1990) fits
the data, but with a normalization twice as large.
  
Before pursuing our analysis, 
it is interesting  to examine the luminosity function by
translating it to a temperature function via the observed $L_X-T$
relation.  In Fig. 1 we show the results of converting various
published luminosity functions by our determination of the local
$L_X-T$ relation (Eq. \ref{eq.LT}).
The luminosity functions considered are from Edge {\etal} (1990),
Henry \& Arnaud (1991) and from the EMSS sample (Gioia {\etal} 1990,
Henry {\etal} 1992).  The temperature data, represented by
the points in the figure, come from the two former sets of authors
(Edge {\etal} 1990 and Henry \& Arnaud 1991).  Thus, only the
EMSS data is entirely independent of the temperature observations.
The converted temperature distribution functions from both 
Henry \& Arnaud (1991) 
and the $z=0.17$ EMSS give slightly flatter functions than the
direct temperature observations of the temperature distribution.
One can see that all of the 
luminosity functions are consistent with the direct determinations 
of the temperature distribution.
Therefore, one would get similar constraints by using luminosity functions
instead of the temperature distribution functions.

In the following, we will compare the theoretical models to 
the observed temperature distribution functions and derive constraints 
on the parameters of the models. 
The free parameters are the index $n$ of the power spectrum
and the bias parameter $b$, both of which appear in the expression for 
the rms mass fluctuations (Eq. \ref{eq.sigma}).
The parameters are derived by chi--square fitting. We fit 
the model to each of the two temperature distribution functions,
which we denote by ESFA and HA for the results of Edge {\etal} (1990) 
and Henry \& Arnaud (1991), respectively.
The differences will be considered as indicative of 
the uncertainties.
We also add an additional point to each data set to represent 
A2163 which is the hottest known cluster. 
The temperature of this cluster was first determined
by Arnaud {\etal} (1992b) who found that is is of the order
of 14 keV with an uncertainty of about 1 keV. 
The importance of this measurement resides in
the precision of the temperature determination. 
Indeed, the uncertainties 
of the cluster temperature  measurements within the samples
of Edge {\etal} (1990) and Henry \& Arnaud (1991) were quite large. 
This has left  open the possibility 
that the existence of such high temperature clusters were not 
real and that they were due to the tail of 
the temperature error distribution.
On the contrary, although the statistical weight of A2163 is small, 
the precision in 
the temperature measurement does give us confidence in the overall shape of 
the distribution functions we used.
We have calculated the number density by assuming 
that A2163 is the 
only such cluster in the Abell survey volume extended to a depth of 
$z \approx 0.3$.  The error bars represented on the data points are 
the $90\%$ confidence limits. Poisson statistics were used in 
the case of HA and A2163.
In Table 1 we give the best fit values of $n$ and $b$ 
for each data set along.  The subscripts
1 and 2 refer to data sets with and without the A2163 
point, respectively.

It is difficult to assess the real nature of the detection 
statistics for clusters and Poisson statistics may not be relevant. 
Accordingly, the error bars should be treated 
with some caution.
In the absence of further knowledge we assume 
that the errors are normally distributed 
and that the given error bars correspond to the 
rms deviation.
We then draw the confidence contours which would contain
68.3\%, 90\% 95.4\% of the normally distributed data 
in the $b-n$ plane (Fig. 2). 
Since the error distribution function is 
unlikely to be gaussian, the actual probability associated 
with these contours cannot be evaluated. However,
we estimated the reliability of the models by checking 
by eye the goodness of fit for different values
of ($b$,$n$) on the 1 $\sigma$ contour.

From the ESFA data set we find:
$$ 1.8 \leq b \leq 1.9  $$
$$ -2.25 \leq n \leq - 1.8 $$
while with the HA data we find:
$$ 1.6 \leq b \leq 1.8  $$
$$ -2.4 \leq n \leq - 1.5 $$
These constraints are similar to those drawn by  
Blanchard \& Silk (1991), Henry \& Arnaud (1991) or more recently
by Bartlett {\etal} (1995).
Because Henry \& Arnaud (1991) use a smaller number of clusters 
spread over a shorter range 
of temperatures, their data set provides less stringent constraints on the 
parameters.
We finally consider as robust the following intervals:
$$ 1.6 \leq b \leq 1.9  $$
$$ -2.4 \leq n \leq - 1.5 $$
 The 
temperature distribution does a good job in constraining both the shape
and the amplitude of the power spectrum.  This 
constraints were derived using clusters with 
temperatures between 2 and 14 keV, so the 
shape of the fluctuation spectrum is actually constrained over  
the range 5 and 10 ${\mbox {\rm h}}^{-1}$Mpc.
One important implication is that the CDM model cannot explain 
the shape of the temperature distribution function:  
its power spectrum is too steep on galaxy cluster scales. 
Instead of the CDM value of $n\approx -1$, the data suggest 
that $n$ is closer to $-2$ over these scales.
This is in agreement with analyses based 
on other methods, for example the power spectrum determination of
Hamilton {\etal} (1991) and Peacock and Dodds (1994). This
conclusion applies independently of the normalization of the 
spectrum. In considering the cosmic microwave
background temperature fluctuations, we may make the additional 
statement that the normalization required by the clusters does 
not conform to the normalization demanded of CDM by the COBE 
measurements: given the CDM spectrum,
the latter favors bias factors of order unity or less.

Our conclusions mainly rely on the validity of the  
temperature--mass relation. It should be emphasized that any
error in this relation enters the exponential of the Gaussian in the 
mass function. As discussed above, the $T-M$ relation may be 
derived from the assumptions of
isothermality and hydrostatic equilibrium and has been 
checked further by numerical simulations.  
However, standard mass estimates from hydrostatic equilibrium may 
underestimate the actual masses of clusters (Balland \&  Blanchard 1996).
  One may make several remarks here concerning future studies
on the relation between the state of the ICM and the underlying 
dark matter.  One will eventually be able to measure the temperature
profile of the gas using the spatially resolved spectroscopic 
data of XMM and AXAF.  For the present, one may attempt to constrain
this profile by combining X--ray images and radio maps of the 
Sunyaev--Zel'dovich effect (assuming sphericity).  Once given a 
temperature profile, the temperature--mass relation may in principle
be deduced only from the assumption of hydrostatic equilibrium.
From our point of view, the most exciting prospect employs the weak 
distortion of gravitationally lensed background 
galaxies to probe the cluster binding mass.  By examining
a sample of clusters with lensing data, X--ray images and
even maps of the Sunyaev--Zel'dovich effect, one can directly
constrain the temperature--mass relation.

\section{Evolution with redshift}

In the above section, we have shown that the local X--ray
data allow one to
fully determine the power spectrum of the density fluctuations.
Oukbir \& Blanchard (1996) used a similar
analysis to constrain the same quantity in the case
of an open universe 
(see also Oukbir \& Blanchard 1992, Viana \& Liddle 1995, 
Eke {\etal} 1996).  Within our framework,
the models are then completely specified
and we can now predict the temperature distribution function of
galaxy clusters at any redshift. 
In principle this is a powerful test of the mean density 
of the universe, since Oukbir \& Blanchard (1996) demonstrated
that the evolution of the comoving number density of
X--ray temperature selected galaxy 
clusters depends solely on $\Omega_0$.
Nevertheless, such information is not yet available and we can only
investigate the evolution of the luminosity function. 
The most straightforward way to achieve this is to use the 
observed $L_X-T$ relation. However, this correlation is 
determined only at low
redshift. On the other hand, the standard scaling relation,
$L_X\propto M^{4/3}(1+z)^{3.5} \propto T^2 (1+z)^{1.5} $,
predicts the evolution with the redshift of the $L_X-T$ relation;
but as we have discussed in Sect. 3.2, this relation 
is not in agreement with local data. 
Actually, there is only little
information concerning the temperature of 
high redshift galaxy clusters and the existing data seem 
to indicate that the 
$L_X-T$ correlation is independent of redshift 
(Henry {\etal}  1994). However, due to the small number of
clusters and to the large error bars on the temperatures,
the uncertainties are quite high and do not lead to robust constraints. 
Investigating the possible evolution of the $L_X-T$ relation, 
Oukbir \& Blanchard (1996) determined the parameters which
best fit the observed redshift distribution of the EMSS
clusters (Gioia \& Luppino 1994). In the case of the $\Omega_0=1$
universe, they found that a non--evolving $L_X-T$ 
relation is in acceptable agreement with the observations, although  
a slight positive evolution $L_X = L_0 (1+z)$ better fits the data
(here, $L_0$ is the luminosity that a cluster of given
temperature would have at $z=0$ according to the local $L_X-T$
correlation). This latter relation is the one we
will use in the following.

The observed evolution of the X--ray cluster population 
has been investigated and discussed in detail
(Gioia {\etal} 1990, Edge {\etal} 1990, Henry {\etal} 1992,
Luppino \& Gioia 1995). 
The situation, however, is not very clear: the first results 
suggested a strong negative evolution, in the sense 
that for a given luminosity, 
fewer clusters were observed at high redshifts.
On the other hand, Ebeling \etal (1995) claim that previous investigations
were undermined by non--uniform selection procedures 
and they found no convincing evidence for any evolution
within a sample of X--ray selected intermediate redshift 
ROSAT clusters, up to $z \sim 0.3$. In fact, 
due to the extended nature of these objects, the 
interpretation of an X--ray selected cluster sample is not 
straightforward: apparent fluxes have to be corrected
by a factor which depends on the assumed geometry
of the source. 
This procedure has been used both by Gioia {\etal} (1990) and  
Henry {\etal} (1992), as well as by Luppino \& Gioia (1995). 
The correction is very large for 
low redshift clusters and becomes moderate at higher redshifts.
For instance, the mean correction factor used by Gioia \& Luppino (1994)
is 7; it could be as high as 15 for clusters with 
redshifts smaller than 0.15, but is less than 1.5 in the highest redshift bins.
It seems therefore possible that a moderate systematic error in this
correction could alter the inferred luminosity function. 

It is interesting to compare our best fitting model with
the observed luminosity function at high redhifts. 
This is presented in Fig. 3, where the luminosity 
function has been computed assuming the above mentioned
evolutionary law for the $L_X-T$ relation.
The most impressive aspect of the observations 
is the fast apparent evolution of the slope of 
the observed luminosity function 
over the moderate redshift range from $\overline{z} = 0.17$ to 
$\overline{z} = 0.33$, which is not reproduced by the models.  
This is manifest by the fact that 
the model curve at $\overline{z} = 0.17$ is already steeper than
the data at this redshift; we have already noted this earlier in
our discussion of Fig. 1.  However, the models remains 
consistent with the data, when the uncertainty are taken into account. 
It is also interesting to note that the data from 
higher redshifts, shown in the inset, demonstrate 
similar or less evolution
than the models; as pointed out by Gioia \& Luppino (1994),
the difference between  $\overline{z} = 0.33$ 
and $\overline{z} = 0.66-0.8$ 
is consistent with no--evolution.  
From all of this,
it seems reasonable to us to conclude that the observations are globally
consistent with a moderate negative evolution, and that this 
evolution is weaker than previously estimated.

\section{The X--ray background and the X--ray counts}

Two interesting probes of cluster evolution are 
the X--ray source counts and the contribution of 
clusters to the X--ray background.

In order to compute the contribution of clusters to the X--ray counts,
we must assign a luminosity to each mass in the 
PS mass function. 
According to Oukbir \& Blanchard (1996), we use the locally 
observed $L_X-T$
correlation and we assume that it evolves such as to best fit the
EMSS cluster redshift distribution (Gioia \& Luppino 1994).
As we mentioned in the previous section, in the case of the $\Omega_0=1$
universe, a non--evolving $L_X-T$
relation is in acceptable agreement with the observations, although
a slight positive evolution $L_X = L_0 (1+z)$ better fits the data.
In the case of an open  universe,
a strong negative evolution is needed
to reproduce the same data, and $L_X = L_0 (1+z)^{-2.3}$.
These latter relations are the one we
will use in the following.

As the models are forced to match
the Einstein redshift distribution as well as the local 
data, we do not expect a significant difference in 
the predicted quantities between the two models.

In Fig. 4, we show the expected
$\mbox{\rm log}\,N(>S)- \mbox{\rm log}\,S$ in the energy band 0.5--2 keV. 
The triangle at $2\times 10^{-12} \mbox{\rm erg}\,\mbox{\rm s}^{-1}
\mbox{\rm cm}^{-2}$
comes from the ROSAT cluster number counts in the northern sky
(Burg {\etal}  1994), and the arrow at  
$10^{-14}\mbox{\rm erg}\,\mbox{\rm s}^{-1}\mbox{\rm cm}^{-2}$ is a lower limit
inferred by Rosati {\etal}  (1995) from deep ROSAT PSPC observations. 
The solid and the dashed lines are the predicted cluster 
counts from our models in the case of the $\Omega_0=1$ and
$\Omega_0=0.2$ models respectively. The thick lines are 
computed assuming the evolution of the $L_X-T$ correlation 
which best fits the EMSS redshift
distribution, whereas the thin line correponds to 
a non--evolving $L_X-T$ relation. 
Although the models are slightly above
the observed number counts at low fluxes,
our self--consistent modeling leads to predicted number counts
which are in agreement with the data, and as expected,
the flat and the open case are then almost identical.
The strong negative evolution that is necessary in open models 
is again emphasized:
if one assumes a non--evolving $L_X-T$ correlation
at high redshifts in the $\Omega_0=0.2$ universe, then one
overproduces the number of expected clusters
at low fluxes by a factor close to five. Although the
counts at $10^{-14}\mbox{\rm erg}\,\mbox{\rm s}^{-1}\mbox{\rm cm}^{-2}$
could constitute a lower limit, it is unlikely that they
were underestimated by such a large factor.

With our approach we can also estimate the contribution of clusters 
to the X--ray background.

The contribution of X--ray clusters to the XRB 
has already been
discussed at length in the literature. From the X--ray luminosity 
function of Abell clusters, McKee {\etal}  (1980) have put an upper limit
of $5\%$ to this contribution in the 2--10 keV band. 
Piccinotti {\etal}  (1982) have used a complete X--ray survey of the 
HEAO experiment to derive 
a similar result for the same energy band. At energies less than 1 keV, 
Schaeffer \& Silk (1988) derived a contribution as high as $50\%$ 
coming from small objects with large redshifts. However, this was 
based on the self--similar model. Blanchard {\etal}  (1992b)
investigated the contribution of X--ray clusters to the XRB
within the framework of different cosmological models. Using parameters
in the luminosity--mass relation which reproduce the local luminosity 
function, they found a contribution of approximately $10\%$ in the 
2--10 keV band. Burg {\etal}  (1993)
have also estimated this contribution, 
but do not attempt to reproduce the locally observed quantities.
Figure 5 shows our calculation of 
the cluster contribution to the X--ray 
background in the energy range 0.07--10 keV.
The two crossing, thin, solid lines correspond to the region 
where the ROSAT background lies (Hasinger 1992). 
The two parallel, solid lines are power laws with energy index of $-0.4$
and two different normalizations at $1\,\mbox{\rm keV}$:
$8\,\mbox{\rm keV}\,\mbox{\rm cm}^{-2}\mbox{\rm s}^{-1}
\mbox{\rm sr}^{-1}{\mbox{\rm keV}}^{-1}$,
as determined from the HEAO spectrum by Marshall {\etal}  (1980),
and $11\,\mbox{\rm keV}\,\mbox{\rm cm}^{-2}\mbox{\rm s}^{-1}
\mbox{\rm sr}^{-1}\mbox{\rm keV}^{-1}$
from the Wisconsin results of McCammon {\etal}  (1983).
The triangles are the values inferred by Wu {\etal}  (1991).
The solid and the dashed lines are the predicted cluster counts
from our models in the case of the $\Omega_0=1$ and $\Omega_0=0.2$
models, respectively. As in Fig. 4, the thick lines
are computed assuming the evolution of the $L_X-T$ correlation 
which best fits the EMSS redshift distribution, whereas the thin
lines correspond to the case of the non--evolving $L_X-T$ correlation.
As for the contribution of clusters to the X--ray counts, 
if the self--consistent modeling is used, then  
the contribution of clusters to the X--ray background is very
similar in the case of the critical and open models.
In the 1 -- 2 keV band, this contribution is about 10\%.
If one considers the hypothesis of a non--evolving $L_X-T$ correlation
in the case of the $\Omega_0=0.2$ model (recall that this model
does not fit the redshift distribution of the EMSS data),
the contribution in the same band reaches 30\%, which 
is still lower than the boundary which is allowed by considering
that approximately 50\% of the background in the 1 -- 2 keV band
is already resolved by point sources (Hasinger 1992). Nevertheless,
Barcons \etal (1994) showed from  fluctuation analysis 
that the extrapolated counts of the present known point sources could 
explain 90\% of the background, but there still remains an unidentified 
component. In this context, a contribution of 10\% is far from negligible.
This is especially true since clusters are 
extended objects and they probably escape detection
by the point source detection
algorithms routinely employed (Blanchard {\etal} 1992b).

We conclude that the self--consistent modeling of galaxy
clusters is in agreement with the observed X--ray cluster
counts, and, contrary to other
claims (Evrard \& Henry 1991, Burg {\etal}  1993), that
the X--ray background 
does not provide us with stringent constraints on the power
spectrum or the evolution of X--ray properties of galaxy clusters.   

\section{Non--Gaussian fluctuations}

In all of the above analysis, we have assumed that the fluctuations 
were Gaussian. This was implicit when we adopted for the function 
$F$ in Eq. \ref{eq.MF}
the PS formula, which is known to 
reproduce the results of numerical simulations in the case of 
Gaussian fluctuations. 
Notice, however, that the appearance of the 
exponential in this formula is not trivial and should 
be considered as fortuitous since the function $F$ is an extremeley complex 
quantity to evaluate, and that this  has not yet been achieved
even in the Gaussian case.
It is not our goal to investigate any
specific case of non--Gaussian fluctuations, since there is
little physical motivation for any specific model.
We would rather like to point out
some differences that may result in such a case.  For non--Gaussian 
fluctuations it is still possible to write the mass function as :
$$ \int_{M}^{\infty} m\Phi(m)dm = \rho \int_{\nu_{\mbox {\tiny NG}}}^
{\infty}F_{\mbox {\tiny NG}}(\nu)d\nu,$$ 
with
$$\nu_{\mbox {\tiny NG}} = \frac{\delta_c}
{\sigma_{\mbox{\tiny NG}}(M)}.$$
However, the function $F_{\mbox {\tiny NG}}$
is now arbitrary. It is then rather simple to show 
that the mass function will mimic a Gaussian fluctuation 
spectrum $\sigma_{\mbox {\tiny G}}$ 
which is related to the non--Gaussian perturbation
spectrum by the following relation:
$$
\int_{\nu_{\mbox {\tiny NG}}}^{\infty}F_{\mbox {\tiny NG}}(\nu)d\nu = 
 \int_{{\delta_c}/{\sigma_{\mbox {\tiny G}}(M)}}^{\infty}F_{\mbox {\tiny G}}
(\nu)d\nu
$$
Because of the arbitrariness of the function $F_{\mbox {\tiny NG}}$, 
it is possible 
to fit the local properties of clusters, whatever the spectrum is, by 
using an appropriate distribution function. As an illustration, 
we have computed the distribution function for which an $n=-1$ spectrum 
would mimic an $n= -2$ spectrum. This is represented on Fig. 6.
As is naively expected, the distribution function presents a tail
towards high $\nu$ which favors the formation of massive clusters.

\section{Conclusions}

The ensemble properties of clusters (and groups) are potentially useful
for the determination of the various ingredients of cosmological models. 
However, in practice there are several problems which must 
be addressed in order to fruitfully use this method.
As clusters are the result of the non--linear collapse of the largest 
fluctuations, one might think that their physical properties would be difficult 
to understand, and that any modeling would suffer from this limitation. 
This is even more 
important when one is trying to account for the evolution of the ensemble 
properties. The optical properties are certainly difficult to model:
as pointed out by Frenk {\etal} (1990), 
projection effects can alter both the optical 
richness and inferred velocity dispersions.  Other problems that 
make the modeling of cluster evolution difficult are 1) evolution of 
the member galaxies; 2) merging; 3) environmental effects likely to 
have played a major -- but yet unclear -- role in the galaxy formation
history.  Therefore any conclusions inferred from optical observations
should be regarded as only tentative. 

On the other hand, X--ray observations appear to provide a more
reliable test of cosmological models because they are much 
less subject to these optical biases.  Nevertheless,
a substantial uncertainty remains in the modeling of cluster X--ray
luminosities because the luminosity depends mainly 
on the cluster core properties. As we have emphasized, the gas temperature 
is better understood from a theoretical point of view, and present day 
data are of a good enough quality to allow reliable modeling. 
We find that the temperature
distribution function indicates a power spectrum index of the
order of $-2.$ and a 
bias parameter of about of $1.7$, in agreement with other
power spectrum determinations.  Notably, this conflicts with the standard CDM
prediction of $n\approx -1$ on cluster scales.
We have also investigated the case of non--gaussian fluctuations:
we show that the local data implied by a  given sprectrum can be 
reproduced by any other 
spectrum, provided that the distribution function of the fluctuations 
is adequatly chosen. Therefore, only specific models can 
be further investigated.
Calculations of the luminosity evolution of cluster 
remains more uncertain, 
mainly because this needs 
to model the cluster core. One way to avoid this problem would 
be to obtain a temperature limited rather than flux limited sample of clusters, 
but this seems rather difficult to achieve. Nevertheless, information
from the redshift distribution of flux limited clusters 
allows one to remove this difficulty.
We find that the redshift distribution of the EMSS clusters can be fitted 
with a moderate positive evolution of the $L_X-T$ (clusters of a given
temperature were brighter in the past) whereas a strong negative 
evolution is needed in the case $\Omega_0 = 0.2$.

Using the best--fit model to the cluster X--ray data, 
we have estimated the predicted cluster number counts as well 
as the contribution of clusters to the X--ray background
in the case $\Omega_0=1$ and $\Omega_0=0.2$.
We confirm previous results: clusters could represent a significant
fraction of the faint sources, and are expected to contribute about
10\% of the X--ray background at energies of the order of a few keV. 
Since our modeling matches the local data as well
as the high redshift observations, there is not a noticeable difference
among the two models in the X--ray counts and the contribution of 
clusters to the X--ray background: within self--consistent modeling, 
these two quantities cannot provide stringent 
constraints on the various models.  
Further observations will help to remove the substantial uncertainty 
in the temperature distribution function and therefore will 
allow a better evaluation 
of the characteristics of the spectrum, while redshift information 
will lead to unambiguous information on the mean density of the 
universe (Oukbir \& Blanchard, 1996).

\newpage
\begin{table}
   \caption{Best--fit parameters}
       \begin{tabular}{ccc}
	 \noalign{\bigskip}
         \hline
         \noalign{\smallskip}
          & $n$& $b$\\
         \noalign{\smallskip}
	 \hline
         \noalign{\smallskip}
	 ESFA$_1$& $-2.02$& $1.84$\\
	 ESFA$_2$& $-2.03$& $1.85$\\	
	 HA$_1$&   $-1.85$& $1.65$\\
	 HA$_2$&   $-1.85$& $1.67$\\
         \noalign{\smallskip}
	 \hline
      \end{tabular}
  \end{table}
\newpage

\leftline{\bf Figures } 
\noindent
\begin{figure}
\psfig {figure=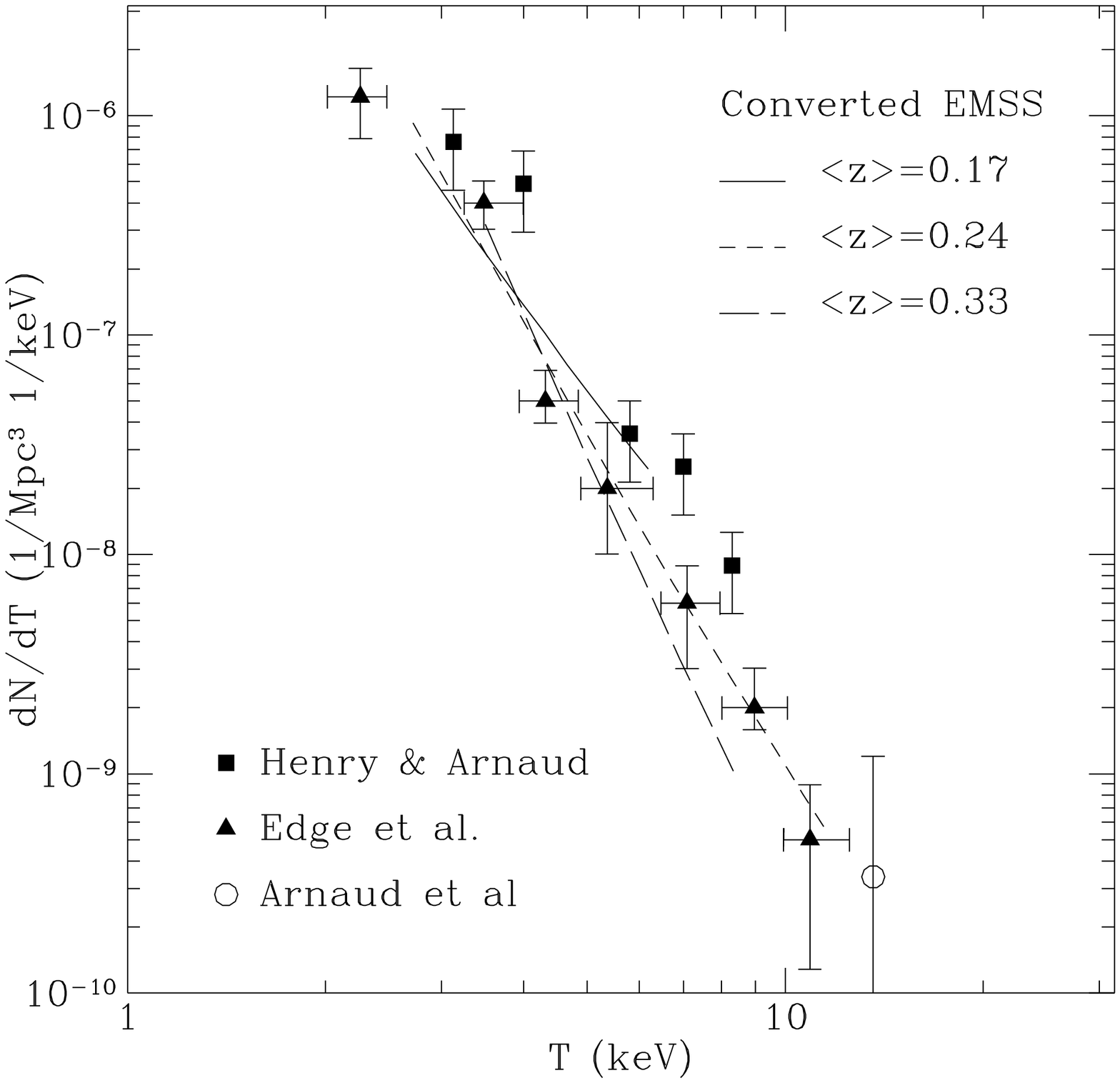,height=12 cm}
\end{figure}

{\bf Fig. 1.}
The two local determinations of the temperature
distribution function, one by Henry and Arnaud (1991), shown as the squares,
and the other by Edge et al. (1990), shown as the triangles, are
compared to the temperature functions deduced from the EMSS  (Gioia
{\etal} 1990, Henry et al. 1992) by application of our $L_X-T$ relation (see text).
In this comparison, the $L_X-T$ relation is assumed NOT to evolve with
redshift.  The different line-types display the results for different
redshifts, as indicated.

\newpage
\noindent
~\vspace*{-2cm}
\begin{figure}
\hbox{\psfig {figure=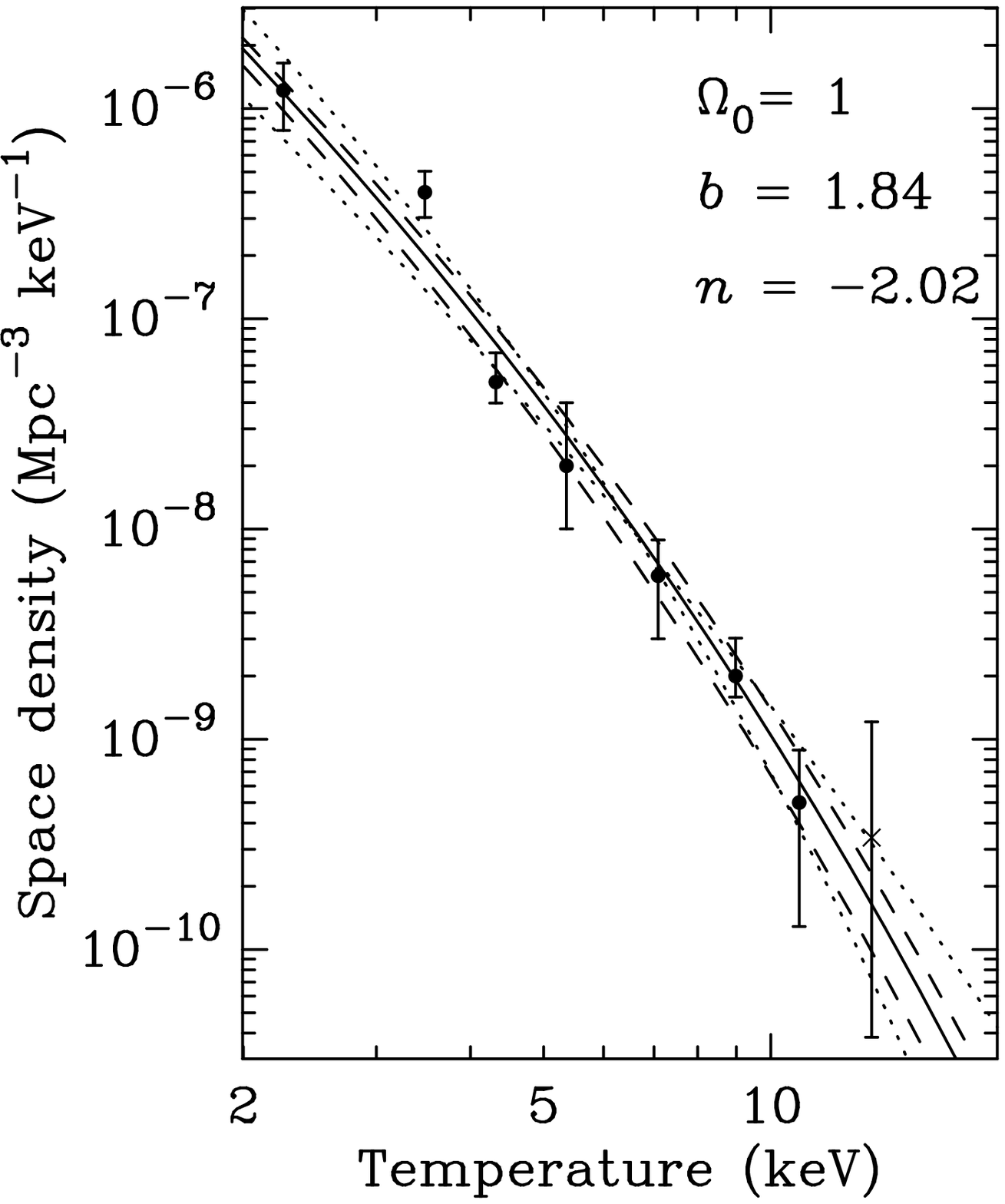,height=12 cm}
\psfig {figure=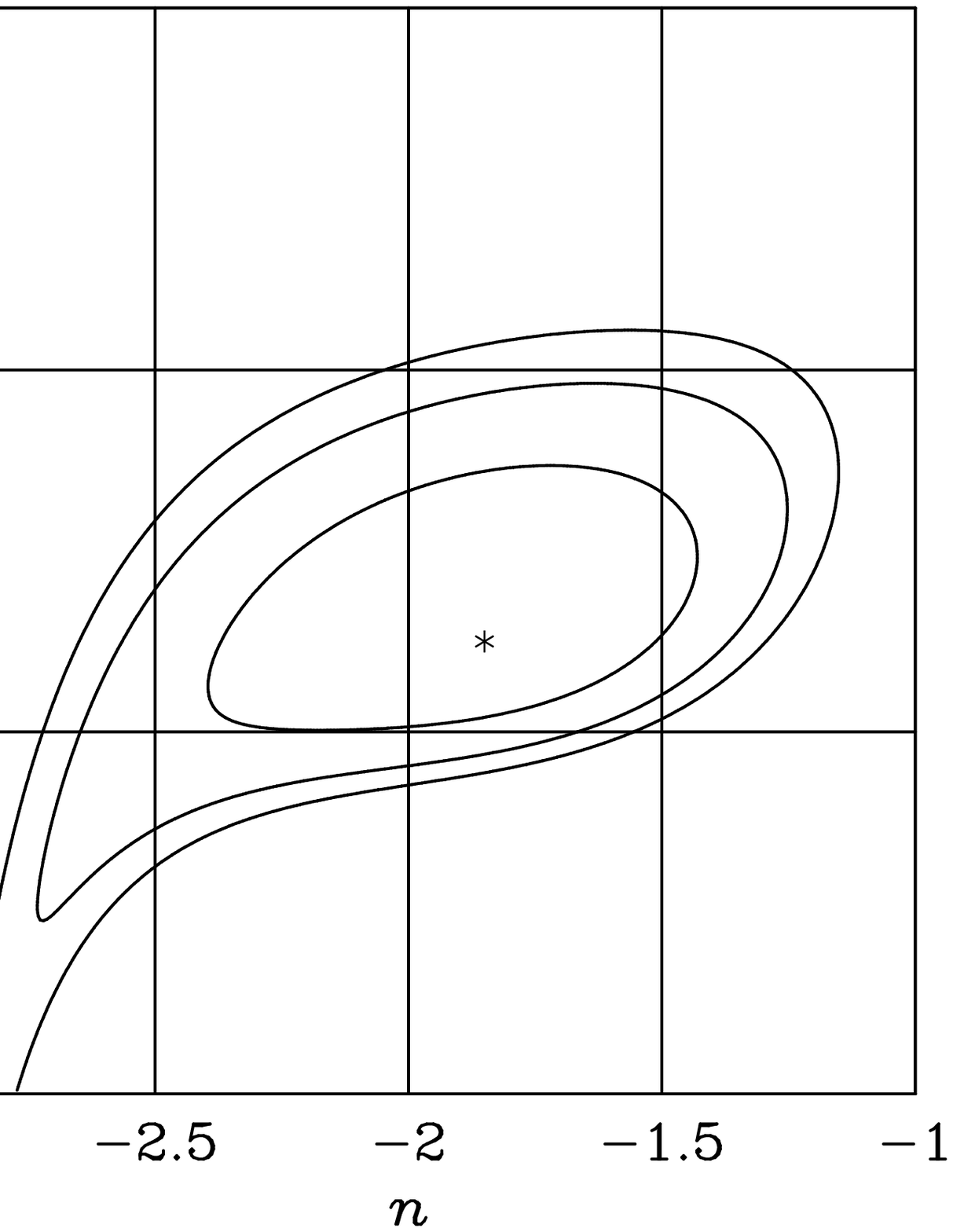,height=12 cm}}
\vspace*{-3.5cm}
\hbox{\psfig {figure=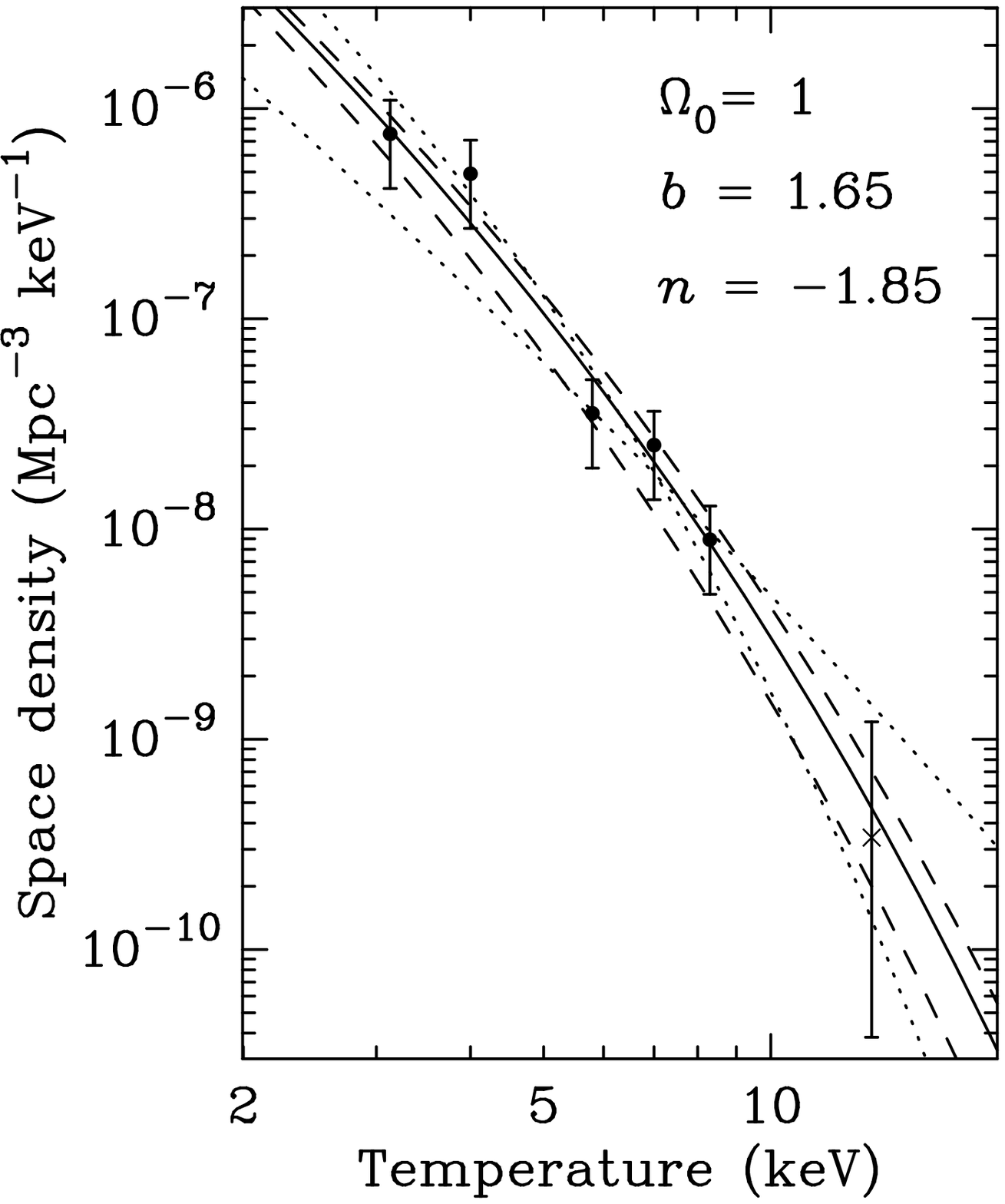,height=12 cm}
\psfig {figure=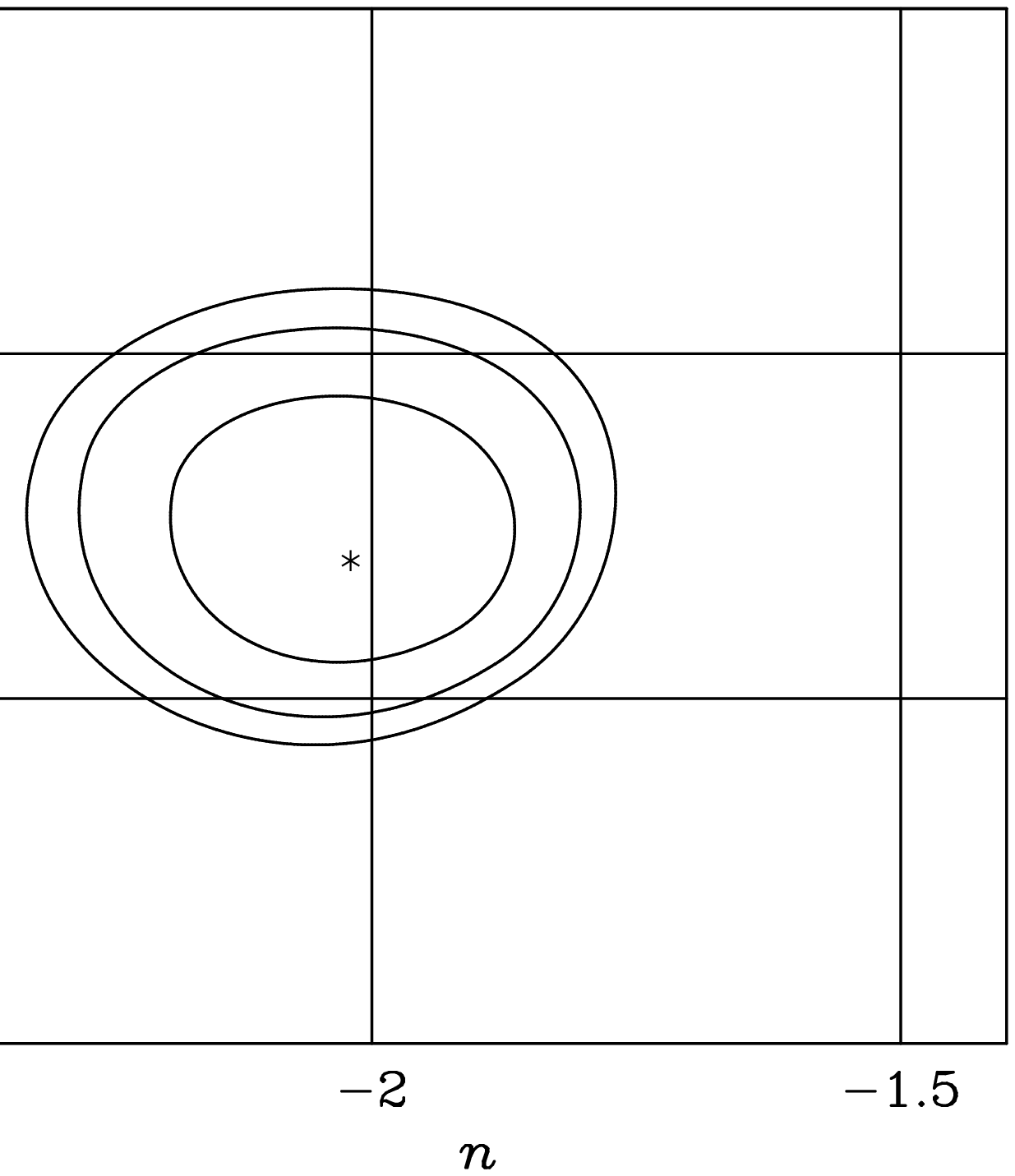,height=12 cm}}
\vspace*{-3.cm}
\end{figure}

{\bf Fig. 2.}
The left hand side column shows observed
temperature distribution functions (points) with the corresponding 
best fitting theoretical functions (solid lines). 
From top to bottom: (a) Henry \& Arnaud (1991) data. 
(b) Edge {\etal} (1990) data.
Temperature distribution functions fitting the data 
at 1 $\sigma$ level are also represented :
(a) The dashed lines represent $b=1.6$ and 1.75 in the case $n=-1.85$
and  the dotted lines represent $n=-2.4$ and $n=-1.5$ in the case $b=1.65$.
(b) The dashed lines represent $b=1.8$ and 1.9 in the case $n=-2.02$
and  the dotted lines represent $n=-2.25$ and $n=-1.8$ in the case $b=1.84$.
The right hand side column shows confidence region ellipses 
corresponding to $\Delta\chi^2=2.30,4.61,6.17$. These contours correspond to 
$68.3\%$, $90\%$ and $95.4\%$ respectively,
for normally distributed data.

\newpage
\vspace{1 cm}
\noindent
\begin{figure}
\psfig {figure=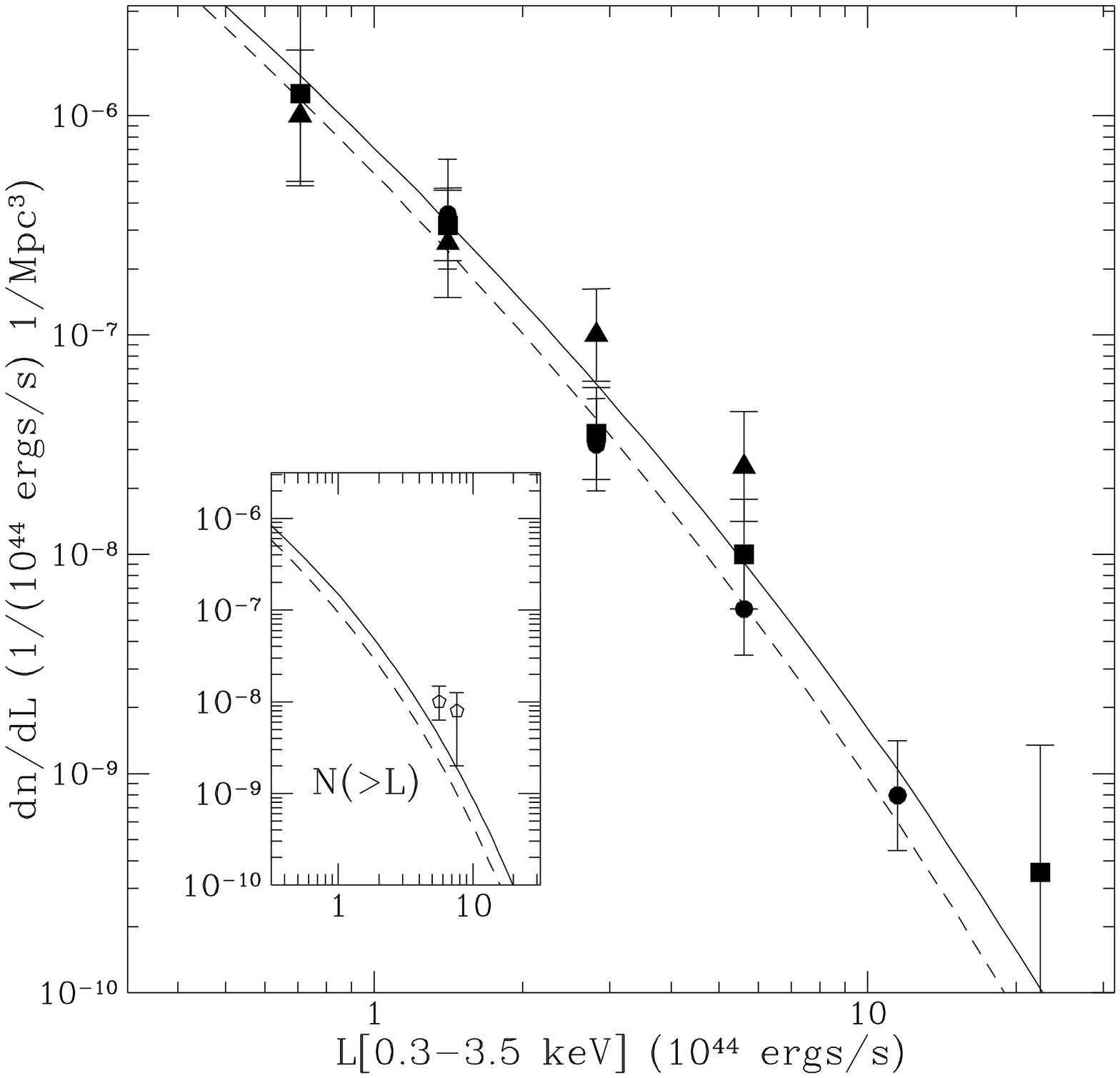,height=12 cm}
\end{figure}

{\bf Fig. 3.}
The X--ray temperature--luminosity function at different
redshifts.  The triangles show the EMSS data at $z=0.17$ and the
squares correspond to $z=0.33$; the solid curve shows the model
redshift--zero luminosity function, while the dashed line shows the
result for a redshift of 0.33.  The model has been normalized to
the local temperature function, and the luminosity functions
have been constructed by application of $L\sim T^3 (1+z)$ to this
local temperature function.  The small inset shows the
integrated number of clusters as a function of in--band
luminosity.  The slightly higher data point corresponds to
a redshift of 0.66 and the lower point to a redshift of 0.8,
both given by Luppino \& Gioia (1995).  The solid and dashed
lines show the corresponding model predictions for $z=0.66$ and
$z=0.8$, respectively.

\newpage
\vspace{1 cm}
\noindent
\begin{figure}
\psfig {figure=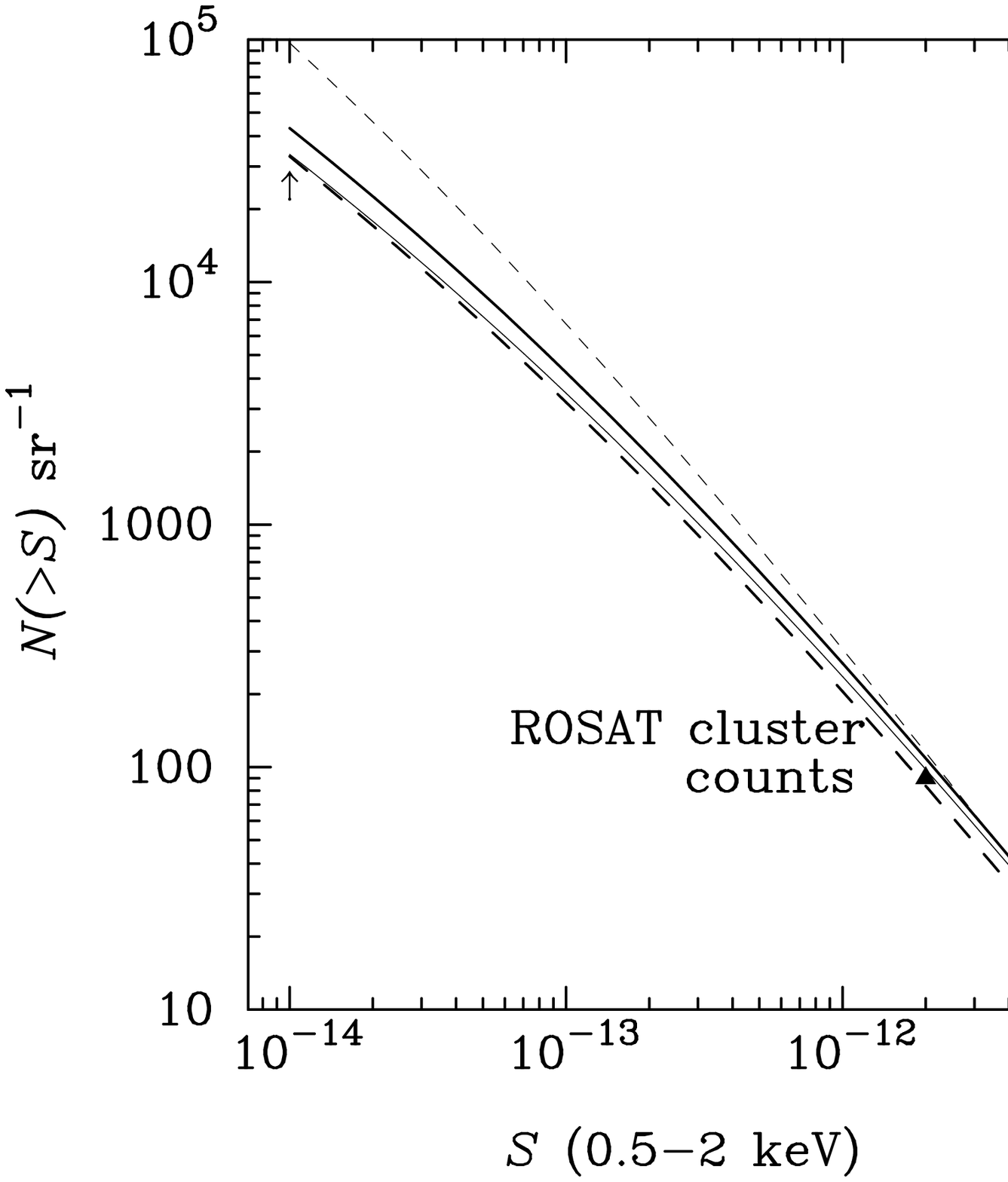,height=12 cm}
\end{figure}

{\bf Fig. 4.}
The galaxy cluster number counts in the 0.5 --2 keV energy band. 
The triangle 
comes from the ROSAT cluster number counts in the northern sky
(Burg {\etal}  1994), and the arrow is a lower limit
inferred by Rosati {\etal} (1995) from deep ROSAT PSPC observations. 
The solid and the dashed lines are the predicted cluster 
counts from our models in the case of the $\Omega_0=1$ and
$\Omega_0=0.2$ models respectively. The thick lines are 
computed assuming that $L_X=L_0(1+z)$ in the case $\Omega_0=1$
and $L_X=L_0(1+z)^{-2.3}$ in the case $\Omega_0=0.2$.
The thin line correponds to 
a non--evolving $L_X-T$ relation. 

\newpage
\vspace{1 cm}
\noindent
\begin{figure}
\psfig {figure=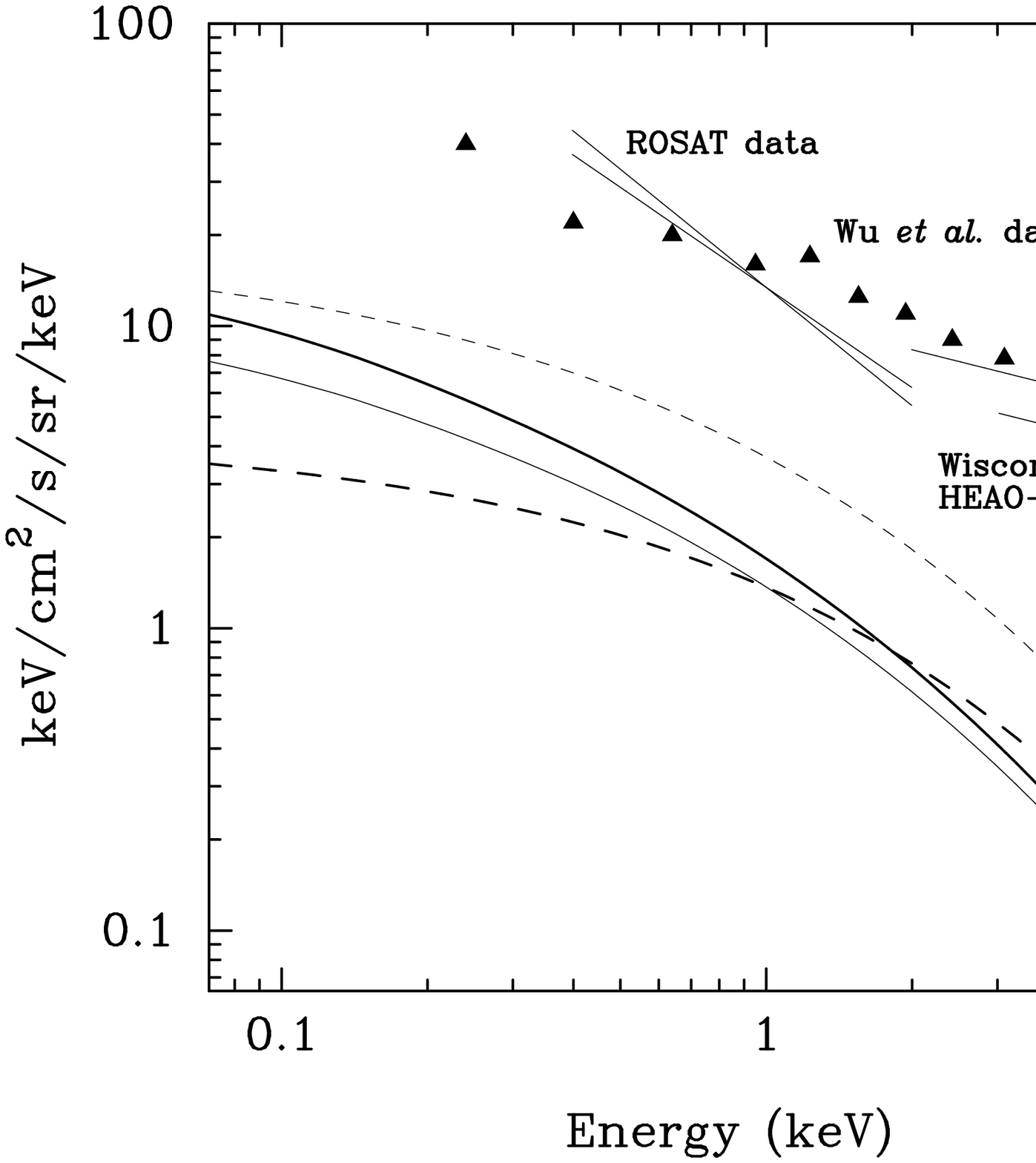,height=12 cm}
\end{figure}

{\bf Fig. 5.}
Contribution of galaxy clusters 
to the XRB in the case of the critical
model (solid lines) and in the case of the open model (dashed lines).
As in Fig. 4, the thick lines are 
computed assuming that $L_X=L_0(1+z)$ in the case $\Omega_0=1$
and $L_X=L_0(1+z)^{-2.3}$ in the case $\Omega_0=0.2$.
The thin solid lines in the range 0.5 -- 2 keV represent the ROSAT background
(Hasinger 1992) and the thin solid lines in the range 2 to 10 keV are
the HEAO 1 background (Marshall {\it et al.} 1980) and the Wisconsin data
(McCammon {\it et al.} 1983) with same power law index and 
different normalisations (see text). The triangles are the values 
inferred by Wu {\it et al.} (1991).

\newpage
\vspace{1 cm}
\noindent
\begin{figure}
\psfig {figure=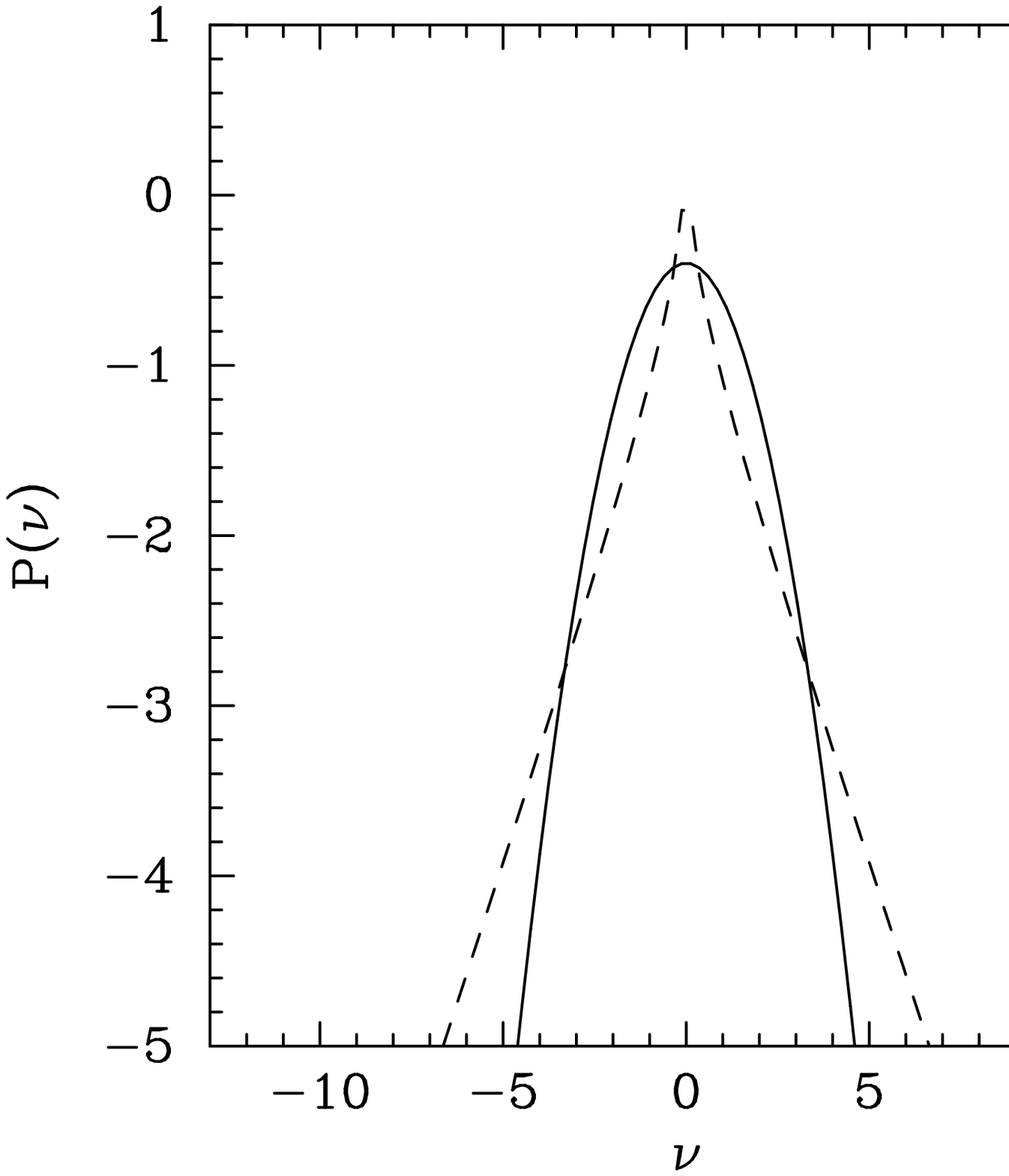,height=12 cm}
\end{figure}

{\bf Fig. 6.}
Probability distribution functions: the Gaussian case (solid line) and the
distribution function for which an $n=-1$ spectrum would mimic 
the mass function obtained with an $n=-2$ spectrum (dashed line)

\end{document}